\documentclass[11pt]{article}
\usepackage{paspconf}
\usepackage{epsf}

\begin{document}

\title{A two-layer $\alpha\omega$ dynamo model, and its implications for 1-D
	dynamos}
\author{Colin B. Roald}
\affil{Center for Space Science and Astrophysics, Varian 302e, Stanford 
	University, Stanford, CA 94305}

\begin{abstract}
I will discuss an attempt at representing an interface dynamo in a
simplified, essentially 1D framework.  The operation of the dynamo is
broken up into two 1D layers, one containing the $\alpha$ effect and the
other containing the $\omega$ effect, and these two layers are allowed
to communicate with each other by the simplest possible representation
of diffusion, an analogue of Newton's law of cooling.  Dynamical
back-reaction of the magnetic field on $\omega$ is included.  I will
show extensive bifurcation diagrams, and contrast them with diagrams I
computed for a comparable purely 1D model.  The bifurcation structure
shows remarkable similarity, but a couple of subtle changes imply
dramatically different physical behaviour for the model.  In
particular, the solar-like dynamo mode found in the 1-layer model is
not stable in the 2-layer version; instead there is an (apparent)
homoclinic bifurcation and a sequence of periodic, quasiperiodic, and
chaotic modes.  I argue that the fragility of these models 
makes them effectively useless as predictors or interpreters of more
complex dynamos.
\end{abstract}


\section{Introduction} %#####################################################

The usual way to apply more computing power to the dynamo problem is
to integrate more and more complex models.  That's not the only way, of
course, because even simple models have multiple free parameters---starting
with the dynamo number itself---whose role in controlling the dynamics 
is rarely investigated in any detail.  The fundamental concern is that
if the behaviour of a system depends sensitively on the values of 
unknown parameters, then it becomes far more difficult to say anything
definite about its interpretation as a solar model.  I have spent some
time mapping the behaviour of some simple 1D dynamos as a function of 
dynamo number and assembling the bifurcation diagrams that show the results.
I do not want to make any strong claims about the particular physical
relevance of these models, but will rather use them as an illustrative
and cautionary example.

This material is presented at greater length in Roald (1998a, b).

\section{Models} %#########################################################

Here we have a standard 1D mean-field $\alpha\omega$ dynamo, with a
dynamical quenching of the shear (Moffatt 1978; Jennings \& Weiss
1991; Roald \& Thomas 1997; Roald 1998a, b),
\begin{eqnarray}
	{\partial A\over\partial t} &=& \mathcal D \cos x\: B 
					+ {\partial^2 A\over \partial x^2}, 
							\label{s1A}\\
	{\partial B\over\partial t} &=& 
		(\sin x +\omega){\partial A\over\partial x} 
					+ {\partial^2 B\over \partial x^2}, 
							\label{s1B}\\
	{\partial \omega\over\partial t} &=& 
		- \left({\partial A\over\partial x}\right)B
				+ \nu{\partial^2 \omega\over\partial x^2},
\end{eqnarray}
where $A$ and $B$ are the toroidal components of the vector potential
and magnetic field, respectively; $x$ is latitude in a quasi-Cartesian
approximation; $\omega=\partial u^\ast/\partial r^\ast$ is the radial
shear, and the third equation controls it by requiring that the
nonlinear terms in the system (i.e., $\omega\partial A/\partial x$ and
$-B\partial A/\partial x$) only exchange energy; and $\nu$ is the
turbulent magnetic Prandtl number.  (The geometry is Cartesian except
that the $\alpha$ and $\omega$ effects have been assigned cosine and
sine dependence.)  Boundary conditions are
\begin{equation}
	A = B = \omega = 0\mbox{ at } x = 0,\pi.
\end{equation}

I compare this system to an essentially similar two-layer, though
still 1D, version of the model that provides a simple representation
of an interface dynamo.  The idea is that the $\alpha$ effect
functions in one layer, just inside the base of the convection zone
(\emph{CZ}), while the $\omega$ effect is concentrated in the shear
layer, which is assumed to be outside and beneath the CZ.  The system
therefore consists of two partial copies of the basic 1D
$\alpha\omega$ dynamo equations (\ref{s1A}--\ref{s1B}), one in the CZ
with an $\alpha$ effect and one in the radiative zone with an $\omega$
effect.  We can connect the two with an analogue of Newton's law of 
cooling, such that the flux between layers is simply proportional to the 
difference between them.\footnote
%-------------------------------------
{This, of course, is a fair approximation only if the dynamo period is much
longer than the diffusion time between layers.  This condition is at best
marginally satisfied, and at worst, quite violated.}
%-------------------------------------
This brings in two additional dimensionless free parameters: the
ratio of the layers' effective diffusivities,
\begin{equation}
	\kappa \equiv \nu_\mathrm{rad}/\nu_\mathrm{conv},
\end{equation}
and the ratio of the shell radius to the separation between layers,
which will enter in the form
\begin{equation}
	\lambda \equiv (R_\mathrm{shell}/d)^2.
\end{equation}

The system of equations resulting from the above construction is:
\begin{eqnarray}
	{\partial a\over\partial t} 
		&=& \mathcal D \cos x\: b + {\partial^2 a\over \partial x^2}
			+\kappa\lambda (A-a), 
							\label{s2a}\\
	{\partial b\over\partial t} 
		&=& \hspace{2em}{\partial^2 b\over \partial x^2} 
			+ \kappa\lambda (B-b) 
							\label{s2b}\\[1ex]
	{\partial A\over\partial t} 
		&=& \hspace{2em}\kappa{\partial^2 A\over \partial x^2} 
			+ \kappa\lambda (a-A), 
							\label{s2A}\\
	{\partial B\over\partial t} 
		&=& (\sin x +\omega){\partial A\over\partial x} 
			+ \kappa {\partial^2 B\over \partial x^2}, 
			+ \kappa\lambda (b-B), 
							\label{s2B}\\[1ex]
	{\partial \omega\over\partial t}  
		&=& - \left({\partial A\over\partial x}\right)B
				+ \nu{\partial^2 \omega\over\partial x^2},
\end{eqnarray}
where $a$ and $b$ describe the $\alpha$ layer, and $A$ and $B$ describe
the $\omega$ layer.

These partial-differential equations were solved by making a spectral
expansion in latitude, then using a standard continuation-method
ordinary-differential-equation solver (\textsc{auto}97).

\section{Results} %#######################################################
\label{Sbd}

\begin{figure}%-----------------------------------------------------------

  \epsfxsize=\textwidth %\epsfysize=0.5\epsfxsize 
  \epsfbox{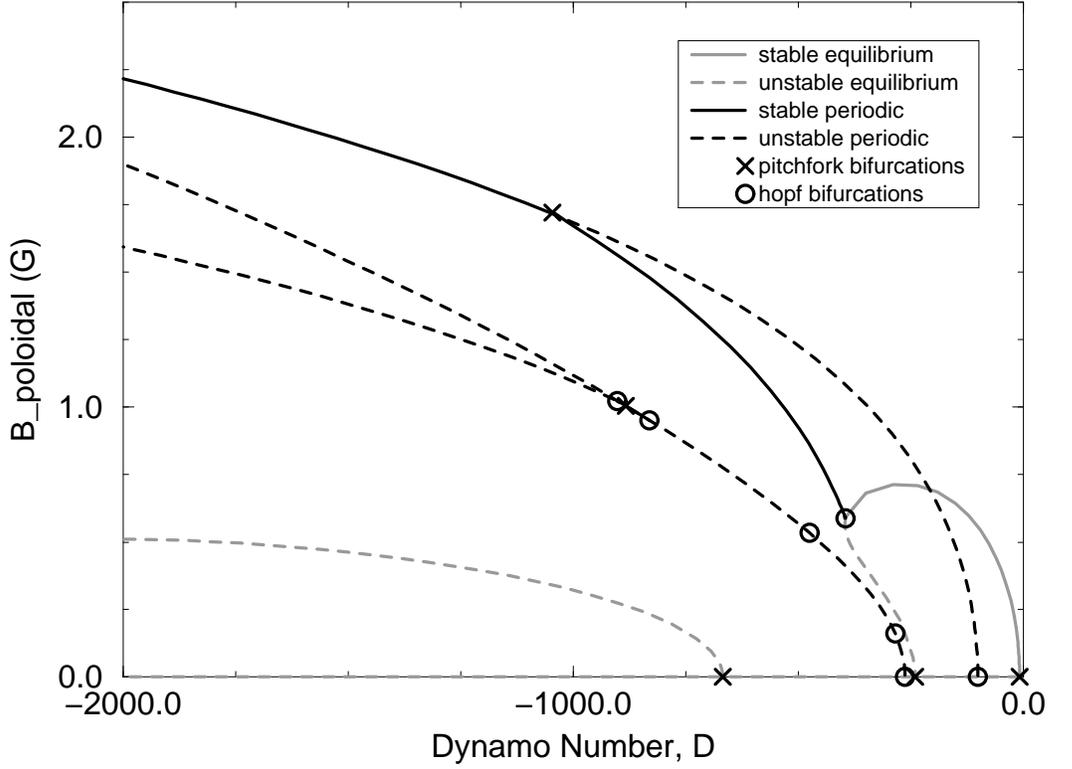}  
  \vspace{-3.25in}  %<--- ugly kludge to correct spacing

  \caption{Bifurcation diagram for the one-layer system,
	$-2000<\mathcal D <0$, $\nu=0.5$, truncation level $N=24$.  See
  	further discussion in \S\ref{Sbd}.}  
  \label{F1}

\end{figure}

\begin{figure}%-----------------------------------------------------------

  \epsfxsize=\textwidth %\epsfysize=0.5\epsfxsize 
  \epsfbox{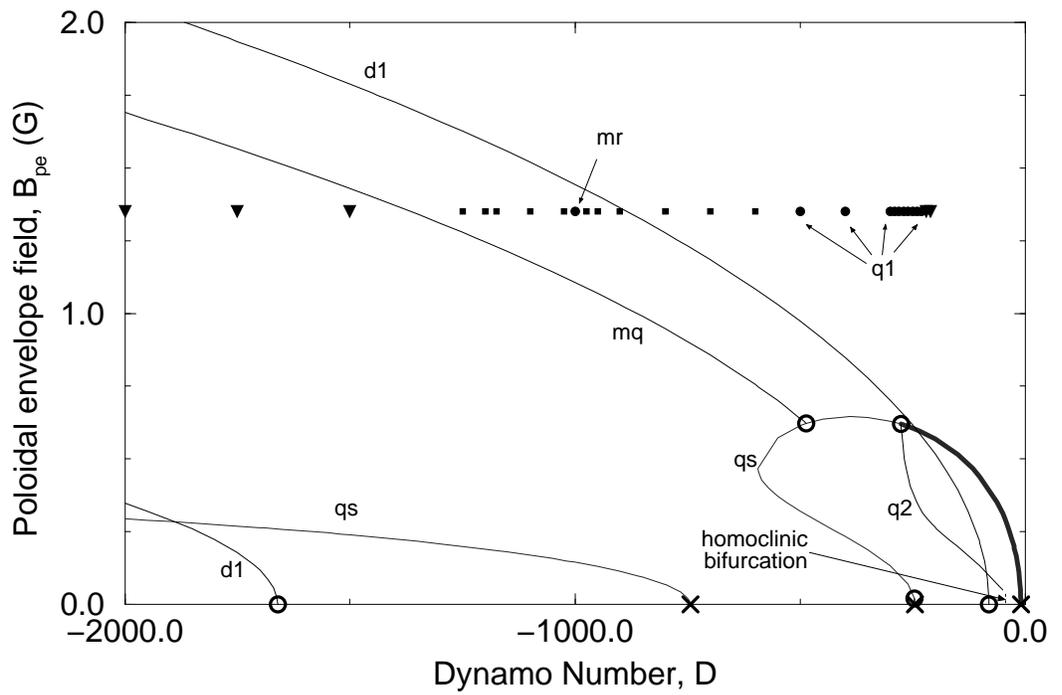}  

  \caption{Bifurcation diagram for the two-layer system, $-2000<\mathcal D <0$,
	$\nu=0.5$, $\kappa=10^{-2}$, $\lambda=600$, truncation level $N=16$.
	See further discussion in \S\ref{Sbd}.}
  \label{F2}

\end{figure}

One set of results from these models is summarised here on a pair of 
bifurcation diagrams (Figures \ref{F1} and \ref{F2}), for the same
magnetic Prandtl number and a comparable range of dynamo numbers
%-------------------------------------------------------------
%-------------------------------------------------------------
%\notetoeditor{I'm afraid I don't know why these postscript figures are
%	rendered with so much unnecessary vertical white space, nor do
%	I really know how to correct it.  I am assuming that the real layout
%	people at the publisher will have a PS editor that can fix the 
%	problem more easily than I can;  if this is not so, please let 
%	me know and I'll try to regenerate them.
%	--- c roald, <roald@stanford.edu>}
%-------------------------------------------------------------------------

Each curve in these diagrams represents a physically distinct solution
of the equations; different solutions are characterised by the
value of their poloidal magnetic field (in the $\alpha$ layer, in the case
of the two-layer model), averaged over latitude and time. Each branch
is labelled with its symmetry, as defined in Roald \& Thomas (1997).  
Bifurcation points are marked.

The isolated points in the lower panels of Figure \ref{F2} (those
plotted with filled circles, boxes, and triangles) show stable
behaviour in parts of parameter space (specifically, for $\mathcal D
\la -200$) where I was unable to get \textsc{auto}97 to lock onto a
solution.  These points were evaluated by time integration and at
lower truncation ($N=8$), and so I have marked them with different
symbols than the \textsc{auto}-computed branches.  Furthermore, these
solutions were classified qualitatively from the appearance of their
Poincar\'e sections, and it should be understood that this technique
is more art than algorithm, particularly for distinguishing between
quasiperiodic and chaotic solutions.  Because transients took a very
long time to decay, I could only evaluate a few of these points for
large negative $\mathcal D$.  And lastly, note that their correct time
average of their poloidal field strength is not determined, so they
are all plotted at a single, arbitrary value.

\section{Discussion} %######################################################

\begin{figure}%-----------------------------------------------------------

  \epsfxsize=3in %\epsfysize=0.5\epsfxsize 
  \epsfbox{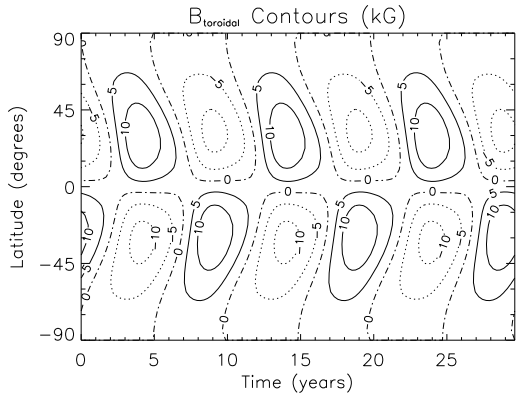}  

  \caption{Contour plots of the variation in time of the toroidal field,
	one-layer model,
	for the stable \emph{mq} periodic branch at $\mathcal D =
	-1011$, $\nu = 0.5$, truncation order $N=24$.}
  \label{Fmq}

\end{figure}

The graphs for the one-layer (Figure \ref{F1}) and two-layer (Figure \ref{F2})
models have many common features, as one would hope for related models.
\begin{itemize}
\item The first bifurcation from the trivial solution---the horizontal 
	axis on the diagrams---is a quadrupole equilibrium, which loops 
	back after a fold.
\item There is another (unstable) quadrupole equilbrium, branching from 
	the trivial solution near $\mathcal D = -700$.
\item A periodic dipole solution branches from the trivial solution near
	$\mathcal D = -100$.
\item A ``mixed quadrupole'' (\emph{mq}) solution branches from the first 
	equilibrium solution shortly before the fold.
\end{itemize}
Similar structure was also found by Jennings \& Weiss (1991) in another
1D model with a different $\omega$-quenching prescription.  The 
mixed quadrupole branch here produces a Sun-like dynamo mode
(Figure \ref{Fmq}).
%-------------------------------------------------------------

On the other hand, we must also note two important differences:
\begin{itemize}
\item the two-layer model (Figure \ref{F2}) does not show an unstable 
	periodic quadrupole solution bifurcating from the trivial solution
	anywhere in the range examined,
\end{itemize}
and critically,
\begin{itemize}
\item the stability of the steady quadrupole solution is destroyed in a 
	subcritical Hopf bifurcation at $\mathcal D = -276$, before the
	supercritical bifurcation to the \emph{mq} solution at $-486$.
\end{itemize}
The consequence of this last is that in the two-layer model, the solar-like
\emph{mq} mode is no longer stable, even though it exists in much the 
same form as in the one-layer version.

\begin{figure}%-----------------------------------------------------------

  \vspace{-1.2in} 
  \epsfxsize=4in\epsfbox{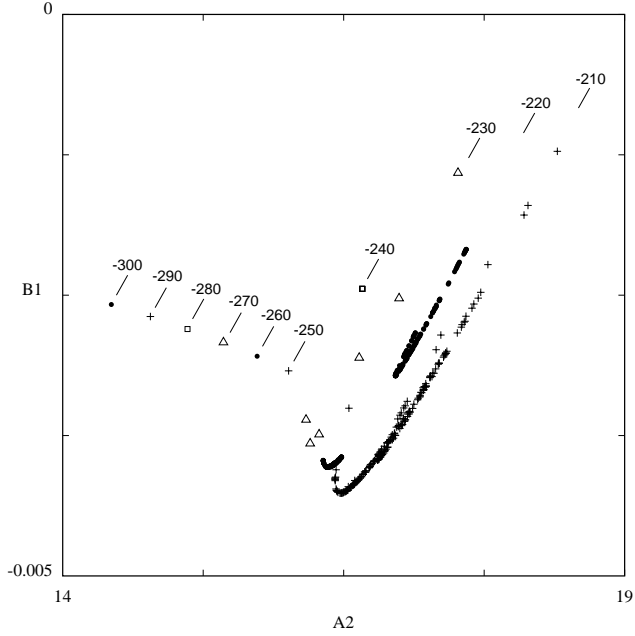}  
  \vspace{-0.7in}

  \caption{Poincar\'e sections for the two-layer system, $\nu = 0.5$, 
	$\kappa=10^{-2}$, $\lambda=600$, truncation $N=8$, taken on the 
	section plane
	$B_3 = 0.005$.  Each section is labelled with its dynamo
	number.  Those for $\mathcal D = -300$ through $-250$ are single
	points, while $\mathcal D = -240$ and $-230$ are period-6 (with the
	six points of $-240$ too close together to be distinguishable
	at this scale).  $\mathcal D = -220$ and $-210$ are apparently 
	chaotic.  All of these solutions have quadrupole symmetry.}
  \label{Fpcar}

\end{figure}

What do we have instead?  The steady quadrupole mode loses stability 
after bifurcation with a quadrupole periodic mode.  This mode tracks
back toward smaller $\mathcal D$, and somewhere around $-60$ appears to 
be destroyed in a homoclinic connection with zero.  At that point, a 
spray of stable unsteady solutions seems to be created, starting with
a chaotic quadrupole.  These do not appear to connect conventionally
with any other solutions, and so I couldn't get the continuation-method
solver to lock on.  Falling back to simple time-integration, we can 
determine the stable behaviour of the system from Poincar\'e sections.
Figure \ref{Fpcar}
%-------------------------------------------------------------
shows solutions for ten different dynamo numbers at the lower end of
the range that shows time-dependent behaviour.  We start with two
chaotic solutions at $\mathcal D = -220$ and $-210$, and progress
through a series of periodic solutions.  This is clearly the crudest
first pass at a serious mathematical investigation of this
system---for example, there are theorems that require some kind of
very complicated dynamics to be going on between the period-6 orbit at
$\mathcal D = -240$ and the period-1 orbit at $-250$---but pursuing it
any farther seems unprofitable from a physical point of view.

\begin{figure}%-----------------------------------------------------------

  \epsfxsize=3in %\epsfysize=0.5\epsfxsize 
  \epsfbox{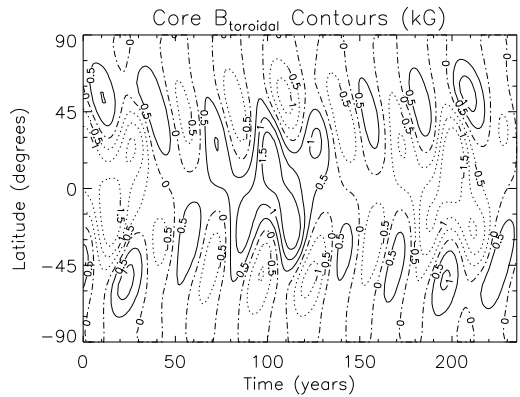}  

  \caption{Contour plots of the variation in time of the shear-layer 
	toroidal field, two-layer model,
	for the stable quadrupole quasiperiodic branch at $\mathcal D =
	-1200$, $\nu = 0.5$, $\kappa=10^{-2}$, $\lambda=600$, 
	truncation order $N=8$.}
  \label{Fqpo}

\end{figure}

These solutions have been marked on the bifurcation diagram, Figure
\ref{F2}.  Figure \ref{Fqpo} shows a typical simulated butterfly 
diagram from the quasiperiodic range.
%-------------------------------------------------------------
None of the computed modes of the two-layer model shows a Sun-like
butterfly.

\section{Conclusions} %######################################################

So, despite a recogniseably similar bifurcation structure, the stable
behaviour of the two-layer model is entirely different from that of the
one-layer version.  This is not in itself a problem, because the two
models do represent different physics.  The concern, however, is that the
only point to studying these simplified, or over-simplified, models is in
the hope that something universal and robust can be identified from them
and applied to our understanding of more sophisticated and computationally
expensive models.  Instead, however, we find almost frightening fragility.

Let me be careful in stating the conclusion here:  I have shown here that
one pair of admittedly unrealistic models are surprisingly sensitive to
the addition of a simple bit of physics.  This does not \emph{directly
predict} anything about the behaviour of more complex models.  What it
does force, however, is the indirect question:  can we be confident that
similar fragility does not occur in other models?  Because if it does, the
models are effectively useless.  Having worked out quite a few bifurcation
diagrams, my personal impression is we would be very lucky to find 
significant details in common between 1D models like these and more 
realistic dynamo models.

\end{document}